\documentclass{WileyMSP-template}
\usepackage{xcolor}
\usepackage{siunitx}
\usepackage{array,multirow,multicol}
\usepackage{colortbl}
\usepackage{booktabs,tabularx,longtable}
\usepackage{rotating}

\begin{document}

\pagestyle{fancy}
\rhead{\includegraphics[width=2.5cm]{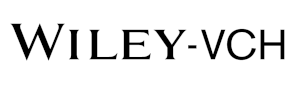}}
\title{Stimulated Forward Brillouin Scattering in Subwavelength Grating Silicon Membranes}

\maketitle


\author{Paula {Nuño Ruano}*},
\author{Jianhao Zhang},
\author{David González-Andrade},
\author{Daniele Melati},
\author{Eric Cassan},
\author{Pavel Cheben},
\author{Laurent Vivien},
\author{Norberto Daniel Lanzillotti-Kimura}, and
\author{Carlos Alonso-Ramos*}


\begin{affiliations}
P. Nuño Ruano, D. González-Andrade, D. Melati, E. Cassan, L. Vivien, N. D. Lanzillotti-Kimura, and C. Alonso-Ramos \\
Centre de Nanosciences et de Nanotechnologies, CNRS, Université Paris-Saclay, Palaiseau 91120, France \\
*Email Address: paula.nuno-ruano@c2n.upsaclay.fr, carlos.ramos@c2n.upsaclay.fr \\

J. Zhang and P. Cheben \\
National Research Council Canada, 1200 Montreal Road, Bldg. M50, Ottawa, Ontario K1A 0R6, Canada \\

P. Nuño Ruano \\
Present address: Photonic Systems Laboratory, École Polytechnique Fédérale de Lausanne, CH-1015 Lausanne, Switzerland \\

D. González-Andrade \\
Present address: Telecommunication Research Institute (TELMA), Universidad de Málaga, E.T.S.I. Telecomunicación, 29010 Málaga, Spain
\end{affiliations}


\begin{abstract}
Brillouin scattering enables efficient and coherent conversion between optical photons and gigahertz-frequency phonons. Integrated circuits that harness this nonlinear interaction have immense potential for signal processing, quantum transduction, and sensing applications. However, achieving strong overlap and tight confinement of optical and mechanical modes in silicon nanophotonic waveguides remains a significant challenge. Here, we propose and demonstrate a novel strategy that enables independent control of optical and mechanical modes in periodically segmented silicon waveguides. Our approach combines two distinct periodic lattices: one with a period shorter than half of the optical wavelength, providing light guiding by metamaterial-induced index contrast, and another that creates a complete phononic bandgap confining acoustic modes. This dual-lattice strategy opens new degrees of freedom to optimize optomechanical confinement and coupling simultaneously. Based on this approach, we experimentally demonstrate remarkably high Brillouin gain of $G_\mathrm{B}=2673$ \unit{\per\W\per\m}, resulting in a Stokes gain of 3 \unit{\dB} and an anti-Stokes loss of 4 \unit{\dB} with 6.4 MHz mechanical linewidth. These results illustrate the potential of subwavelength silicon metamaterials for engineering on-chip optomechanical interactions.
\end{abstract}


\section{Introduction}
Brillouin scattering refers to the inelastic light scattering by acoustic phonons. The stimulated Brillouin scattering (SBS) regime yields the strongest third-order optical nonlinearity for many dielectric materials. Brillouin scattering is currently exploited in optical fibers and integrated devices to implement a wide range of devices, including ultra-narrow linewidth lasers \cite{SubHertz_BSlaser_Gundavarapu2019}, radio-frequency (RF) signal generators \cite{RFsource_Merklein2016} and filters \cite{RFfilter_PPERO_Kittlaus2018}, distributed sensors \cite{DistributedSensorFibre_Murray2022}, optical storage \cite{LightStorage_MerkleinStiller2020}, and quantum applications \cite{QuantumBSwaveguide_Stiller2023}.

SBS has been widely utilized by the optical fiber community \cite{review_fibres_BS}. More recently, on-chip optomechanical micro- and nanostructures have attracted great interest for their ability to enhance and tailor these interactions through a combination of electrostriction and radiation pressure forces \cite{peter_t_rakich_giant_2012}. Silicon-based devices have experienced noteworthy progress since the first experimental observation in 2013 \cite{shin_tailorable_2013} due to the strong optomechanical interactions and compatibility with standard CMOS technology \cite{eggleton_brillouin_2019}. 

Nevertheless, the silicon-on-insulator (SOI) platform faces a significant challenge: the simultaneous confinement of optical and mechanical modes within the waveguide. Strong phonon leakage towards the silica undercladding hinders efficient Brillouin interactions in SOI geometries \cite{wiederhecker_brillouin_2019}. This limitation cab be overcome by using suspended or quasi-suspended structures. Membrane rib waveguides \cite{kittlaus_large_2016}, fully suspended nanowires \cite{laer_net_2015}, pedestal waveguides \cite{van_laer_interaction_2015}, and anti-resonant acoustic waveguide membranes \cite{lei_antiresonant_2024} have proven to be an efficient solution to achieve large Brillouin. However, these structures often show limited mechanical stability or rely on extremely low optical loss to compensate for a moderate Brillouin gain. Periodic waveguides have been proposed to maximize the strength of the Brillouin interactions, achieving optical and mechanical confinement by simultaneous optical and mechanical bandgaps \cite{phoxonicWG_Laude_2011}. However, these phoxonic crystals require large lattice constants to confine GHz mechanical modes, which induce significant optical losses. Therefore, experimental demonstration of optomechanical interactions using these geometries is limited to microcavities \cite{gomisClivia_1DPhoxonic_2014}. Subwavelength grating (SWG) waveguides, characterized by periodicities smaller than half the operating wavelength, exhibit low optical propagation losses and offer exceptional versatility for the design and optimization of photonic devices \cite{cheben_subwavelength_2018, Cheben_MetamaterialsIntegrated_2023}. Theoretical studies suggest that optical and mechanical modes could be confined in SWG waveguides \cite{schmidt_suspendedMIR_2019, zhang_subwavelength_2020, Gonzalez-Andrade_theoryTMBS_2024}, where sections of different thicknesses ensure acoustic confinement based on impedance mismatch. Experimental validation of this approach has not yet been addressed.

In this work, we propose and experimentally demonstrate Brillouin interactions in SWG silicon waveguides. We utilize forward stimulated Brillouin scattering (FBSB), which couples longitudinally propagating photons with transversely propagating phonons \cite{wiederhecker_brillouin_2019}, to design longitudinal and transversal geometries that allow for independent control of photonic and phononic modes. We combine two silicon lattices with different longitudinal and transversal periods (see Figure \ref{fig:geometry}). Near the waveguide core, we use a longitudinal period shorter than half of the wavelength, confining the optical mode by metamaterial effective index contrast \cite{cheben_subwavelength_2018}. At the outermost parts of the waveguide, we implement a square lattice with a period resulting in a total bandgap for the mechanical modes. The proposed geometry, which is fabricated with a single lithography step, yields a significant measured Brillouin effect of up to 1.5 \unit{\dB} of Stokes gain and a 2 \unit{\dB} of anti-Stokes depletion in a 6 \unit{\mm}-long waveguide with 35.5 \unit{\mW} of input power. Additionally, we show the tunability of the SBS resonant frequency between 5 GHz and 7 GHz by varying the transversal waveguide geometry without without negatively affecting performance.  To the best of our knowledge, this is the first experimental demonstration highlighting the potential of subwavelength engineering  in the implementation of optomechanical waveguides.

\section{Design and Simulation}
\begin{figure}
    \centering
    \includegraphics[width=\linewidth]{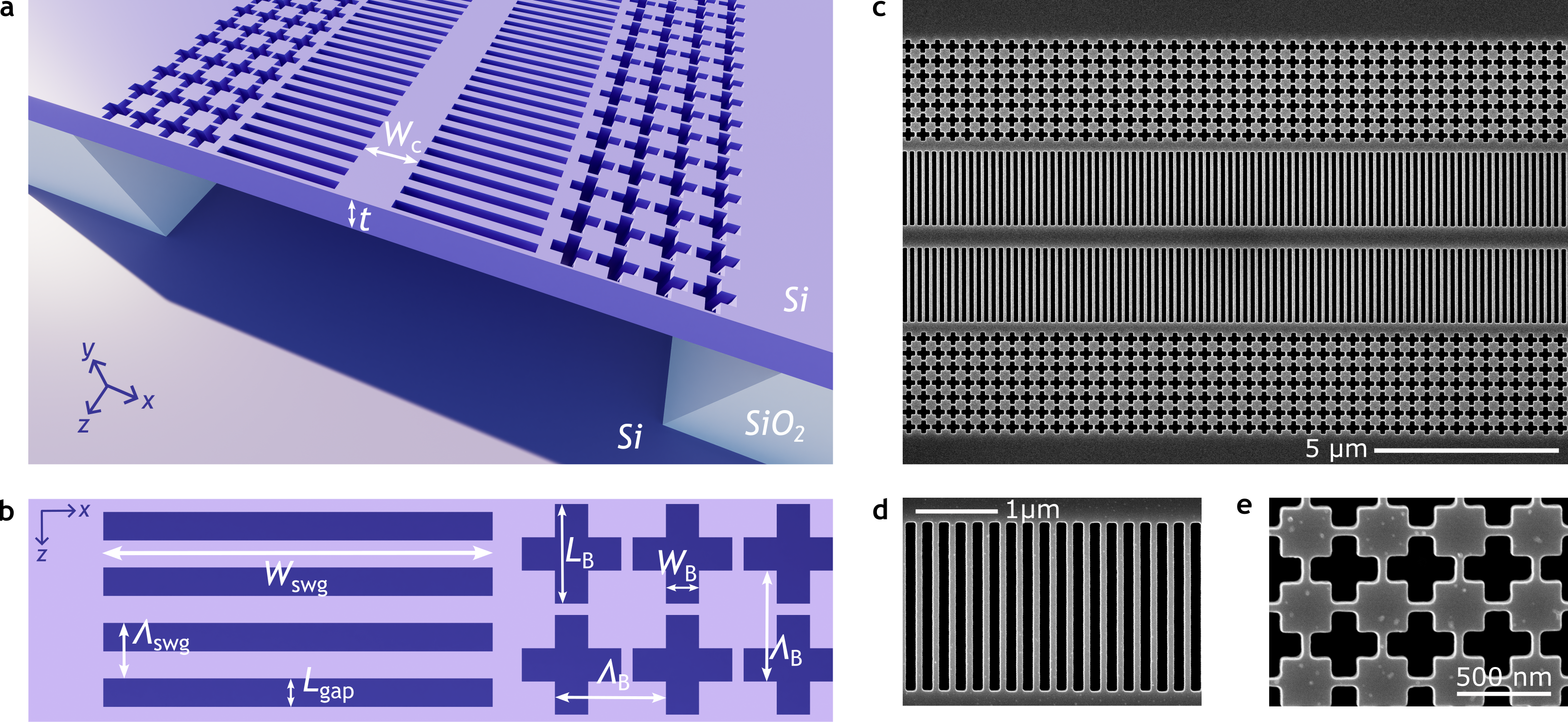}
    \caption{Proposed optomechanical waveguide. \textbf{a} and \textbf{b} Schematic view of the proposed design, with the definition of the geometrical parameters. \textbf{c} to \textbf{e} SEM images of the fabricated waveguide.}
    \label{fig:geometry}
\end{figure}

The optomechanical waveguide, illustrated in Figure \ref{fig:geometry}, consists of a suspended central core surrounded by a cladding made up of a lattice of silicon arms and a phononic crystal barrier. The inner section of the cladding ensures propagation in the subwavelength regime for the fundamental transverse electric (TE) optical mode at 1550 nm wavelength. We engineer the SWG cladding dimensions ($W_\mathrm{swg}, L_\mathrm{swg}$) to minimize the interaction of the optical mode with the phononic crystal, thereby reducing optical propagation losses. The phononic crystal features an air-cross unit cell arranged in a two-dimensional square lattice \cite{PhnC_th_Wang2013}. This design aims to confine a set of mechanical modes by Bragg reflections within the waveguide core and inner cladding. The entire waveguide is fabricated using a single-etch step.

\begin{figure}[hbtp]
    \centering
    \includegraphics[width=0.65\linewidth]{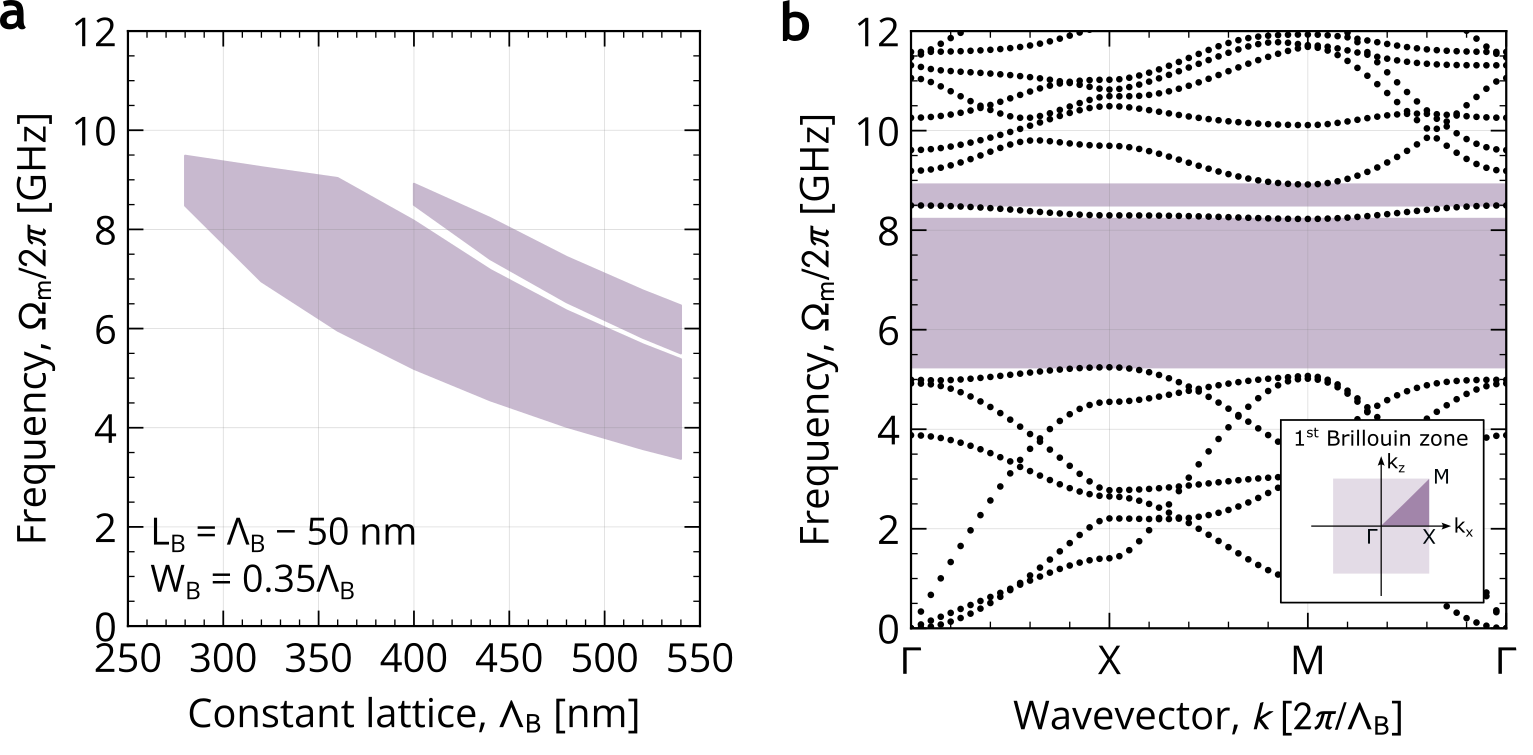}
    \caption{Phononic crystal design. \textbf{a} Bandgap width  as a function of the lattice constant and \textbf{b} full band structure of a phononic crystal with dimensions: $\Lambda_\mathrm{B} = 400$ nm, $L_\mathrm{B} = 350$ nm, $W_\mathrm{B} =  140$ nm. In the inset, a schematic of the first Brillouin zone with the highly symmetrical points in the $k-$space.}
    \label{fig:PhnC}
\end{figure}

To confine adequately the breathing mechanical modes of the central strip waveguide, their frequency must fall within the bandgap of the phononic crystal. In a first approximation, one can picture the waveguide as a Fabry-Pérot cavity for the transversal mechanical resonances. Consequently, the acoustic frequency can be estimated as $\Omega \approx v_\ell/2w$, where $v_\ell \approx 8430$ \unit{\m\per\s} is longitudinal acoustic velocity in bulk silicon and $w$ is the waveguide relevant dimension. The fundamental TE optical mode couples efficiently with the breathing modes in the $x$ direction. For waveguide widths $W_\mathrm{swg}$ ranging from 500 to 700 nm, the mechanical transversal resonances have frequencies between 6 and 8.5 GHz. The selected geometry for the phononic crystal presents a complete bandgap for a lattice constant above 280 \unit{\nm}. However, only periods between 350 and 450 \unit{\nm} exhibit the complete phononic bandgap in the appropriate frequency range (Figure \ref{fig:PhnC}a). These periods are larger than the Bragg period for the TE fundamental mode at 1550 nm ($\Lambda_\mathrm{Bragg} \sim 300$ nm) and, thus, placing the phononic crystal near the waveguide core would induce high propagation losses. In Figure \ref{fig:PhnC}b, we illustrate the entire band structure computed using COMSOL mechanical simulations for a phononic crystal  with a lattice constant of $\Lambda_\mathrm{B}=400$ \unit{\nm} with an air cross with dimensions $L_\mathrm{B} = 350$ \unit{\nm} and $W_\mathrm{B} =  140$ \unit{\nm} (see Figure \ref{fig:geometry}b for the definition of each geometry parameter). 

\begin{figure}
    \centering
    \includegraphics[width=\linewidth]{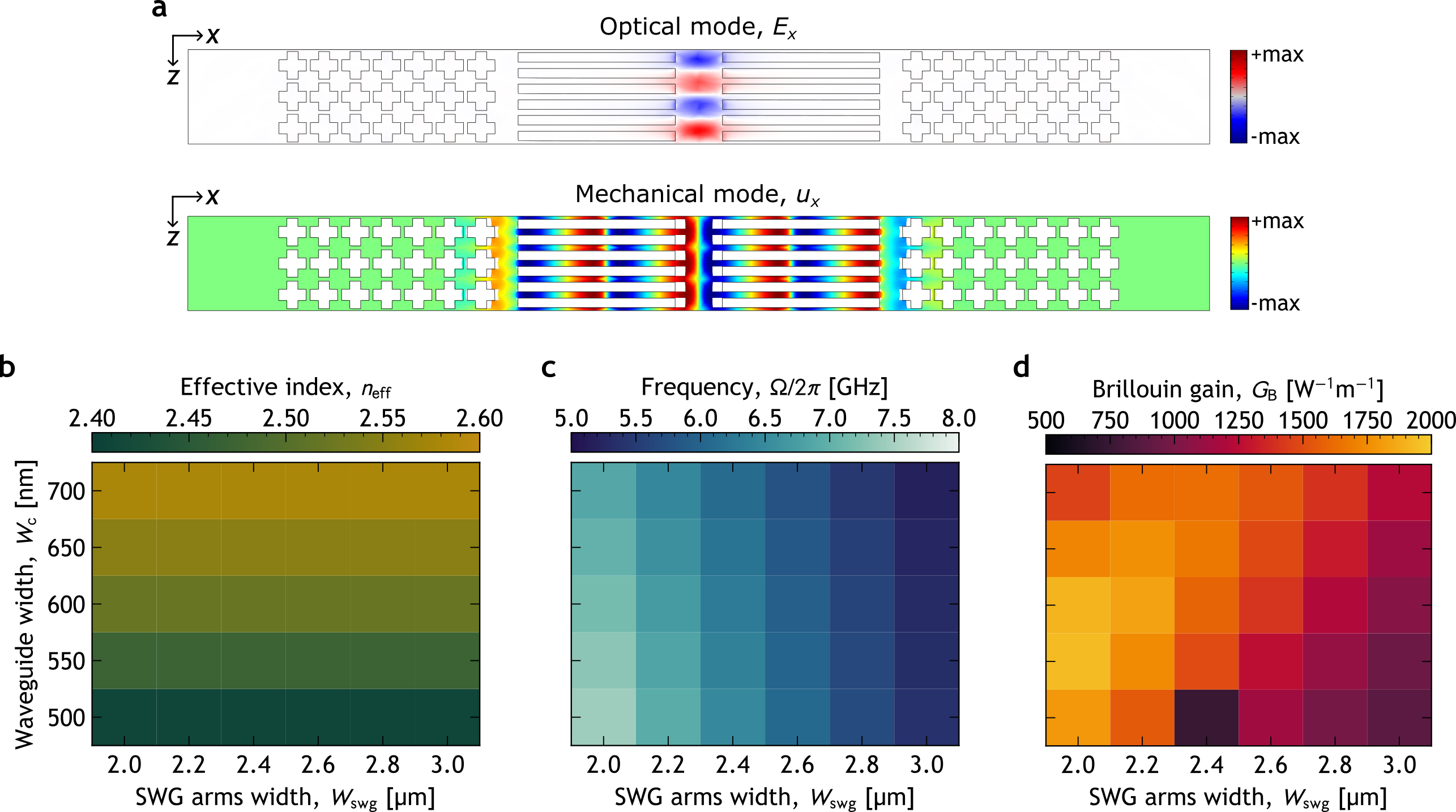}
    \caption{Optomechanical simulations. \textbf{a} fundamental TE optical mode ($E_x$ component) at 1550 \unit{\nm} wavelength (top) and breathing mechanical mode ($u_x$ displacement) at a frequency of 7.184 \unit{\GHz} (down). \textbf{b} Effective mode index for the fundamental TE mode, \textbf{c} mechanical frequency, and \textbf{d} Brillouin gain dependency with the geometry. For the gain calculations, we assume a quality factor of $Q_\mathrm{m}=1000$ and an optical frequency of $\omega_\mathrm{p}/2\pi = 193$ THz.}
    \label{fig:modes_coupling}
\end{figure}

The SWG cladding isolates the optical mode from the phononic crystal. We select a periodicity of 200 \unit{\nm} to ensure better compatibility with the phononic crystal lattice. In Figure \ref{fig:modes_coupling}a, we illustrate the fundamental TE optical mode at a wavelength of 1550 \unit{\nm}, along with the displacement profile for the mechanical breathing mode. The main component, $u_x$, is anti-symmetric with respect to the center of the waveguide core and extend across the entire width of the cavity formed between the two phononic crystals. Several order mechanical modes can exists, characterized by the number of nodes for the mechanical displacement in the subwavelength arms (see Supplementary materials, section D). Only those whose frequency falls within the phononic crystal are well confined and couple efficiently to the light. In Figure \ref{fig:modes_coupling}a, we display the third order mode for reference.

We analyze how the properties of the optical, mechanical, and optomechanical waveguide vary as a function of the widths of the waveguide core ($W_\mathrm{c}$) and the SWG cladding ($W_\mathrm{swg}$). Initially, we observe that the effective index of the optical mode shows little dependence on the width of the SWG cladding, while it is strongly influenced by the width of the waveguide core (Figure \ref{fig:modes_coupling}b). On the other hand, the mechanical frequency varies significantly with changes in both the widths of the waveguide core and the SWG cladding (Figure \ref{fig:modes_coupling}c), consistent with the naive Fabry-Pérot picture. The Brillouin gain coefficient reaches its maximum for a waveguide core width near $W_\mathrm{c} = 550$ nm and SWG cladding width of $W_\mathrm{swg}=2$ \unit{\um}, as shown in Figure \ref{fig:modes_coupling}d, resulting from the interplay between photoelastic and moving boundary effects.

\section{Experimental Results}
We fabricate the device from a standard (100) SOI wafer with a top silicon layer thickness of $t = 220$ \unit{\nm}. The devices are defined using electron-beam lithography and ICP-RIE dry etching. The buried oxide layer is removed using hydrofluoric acid vapor. The optomechanical waveguide is aligned with the $ \langle 110 \rangle$ direction of crystalline silicon, with a total length of 6 mm and a central core width of $W_\mathrm{c} = 600$ \unit{\nm}. The SWG cladding has a period of $\Lambda_\mathrm{swg} = 200$ \unit{\nm}, a gap of $L_\mathrm{gap} = 120$ \unit{\nm}, and a width of $W_\mathrm{swg} = 2$ \unit{\um}, which isolates the optical mode from the phononic crystal. The latter consists of an air-cross unit cell of dimensions $L_\mathrm{B} = 350$ \unit{\nm} and $W_\mathrm{B} =  140$ \unit{\nm} placed along a square lattice of constant $\Lambda_\mathrm{B} = 400$ \unit{\nm}. 

\begin{figure}
    \centering
    \includegraphics[width=\linewidth]{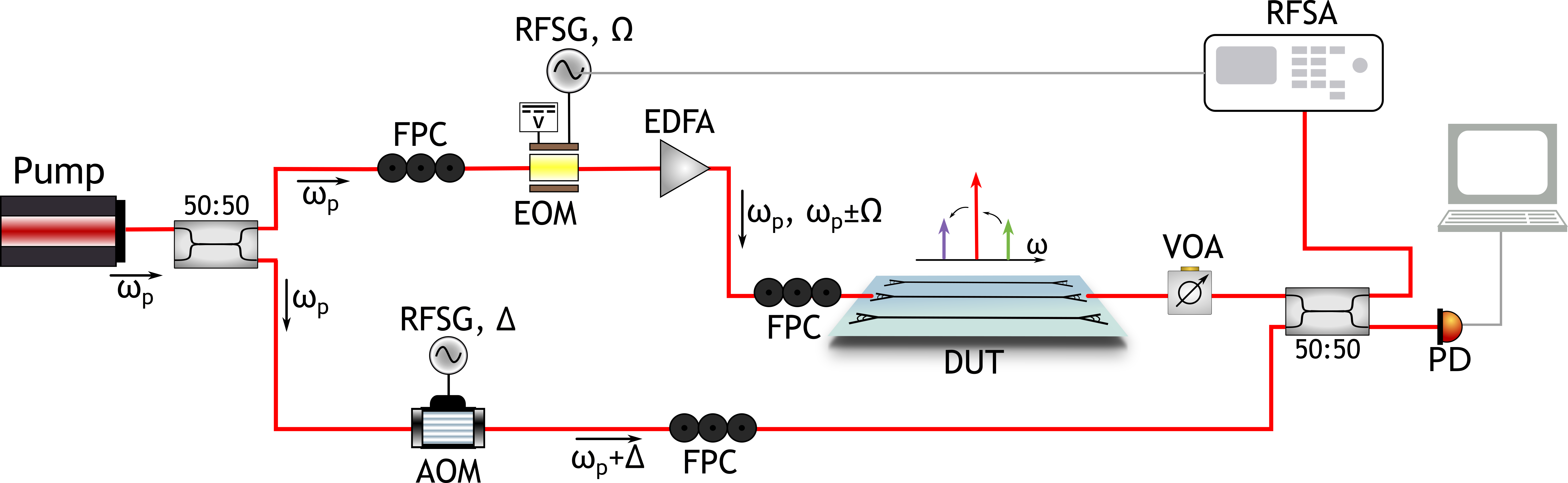}
    \caption{Experimental setup. Two processes of energy transfer take place in the waveguide: from the blue-detuned anti-Stokes sideband $(\omega_\mathrm{as})$ to the pump line $(\omega_\mathrm{p})$, and from the pump line to the red-detuned Stokes sideband $(\omega_\mathrm{s})$. A reference line with up-shifted frequency with respect to the pump $(\omega_\mathrm{p}+\Delta)$ is used to distinguish between both sidelines in the RF spectrum. Notations: AOM, acousto-optic modulator; DUT, device under test, EDFA, erbium-doped fiber amplifier; EOM, electro-optic modulator; FPC, fiber polarization controller; PD, photodetector; RFSA, RF spectrum analyzer; RFSG, RF signal generator;  VOA, variable optical attenuator. In red, we show the optical links and in grey the electrical ones.}
    \label{fig:exp-setup}
\end{figure}

Forward Brillouin scattering in low-dispersive waveguides couples the Stokes and anti-Stokes processes. As a result, spontaneous Brillouin interactions are very inefficient \cite{Noise_dynamics_2016}, and a stimulated process is necessary to obtain measurable Brillouin gain. To analyze the optomechanical response, we conduct a three-tone gain experiment \cite{ACS_Photonics_3toneGain}, as shown in Figure \ref{fig:exp-setup}. We use a narrow-linewidth laser source at 1550 \unit{\nm} wavelength as the pump. The laser is modulated by an electrically driven intensity modulator to generate weak Stokes and anti-Stokes sidebands. These three waves are then injected into the waveguide, where they interact through optomechanical interactions as the modulation frequency $\Omega$ sweeps across the Brillouin resonance. A reference line is added at the device's output by up-shifting the pump frequency by $\Delta = 40$ \unit{\MHz} with an acousto-optic modulator. The Stokes and anti-Stokes intensities are measured by heterodyne detection as an RF beating note using a fast photodetector and a radio-frequency (RF) spectrum analyzer. This experimental configuration can be interpreted as a gain enhancement scenario, where two processes occur simultaneously: energy transfer from the anti-Stokes line to the pump and energy transfer from the pump to the Stokes line. In the subsequent discussions, the terms Brillouin gain or loss will refer to the relative change in intensity of the Stokes and anti-Stokes lines. All experimental measurements are conducted at room temperature and ambient pressure.

\begin{figure}
    \centering
    \includegraphics[width=\linewidth]{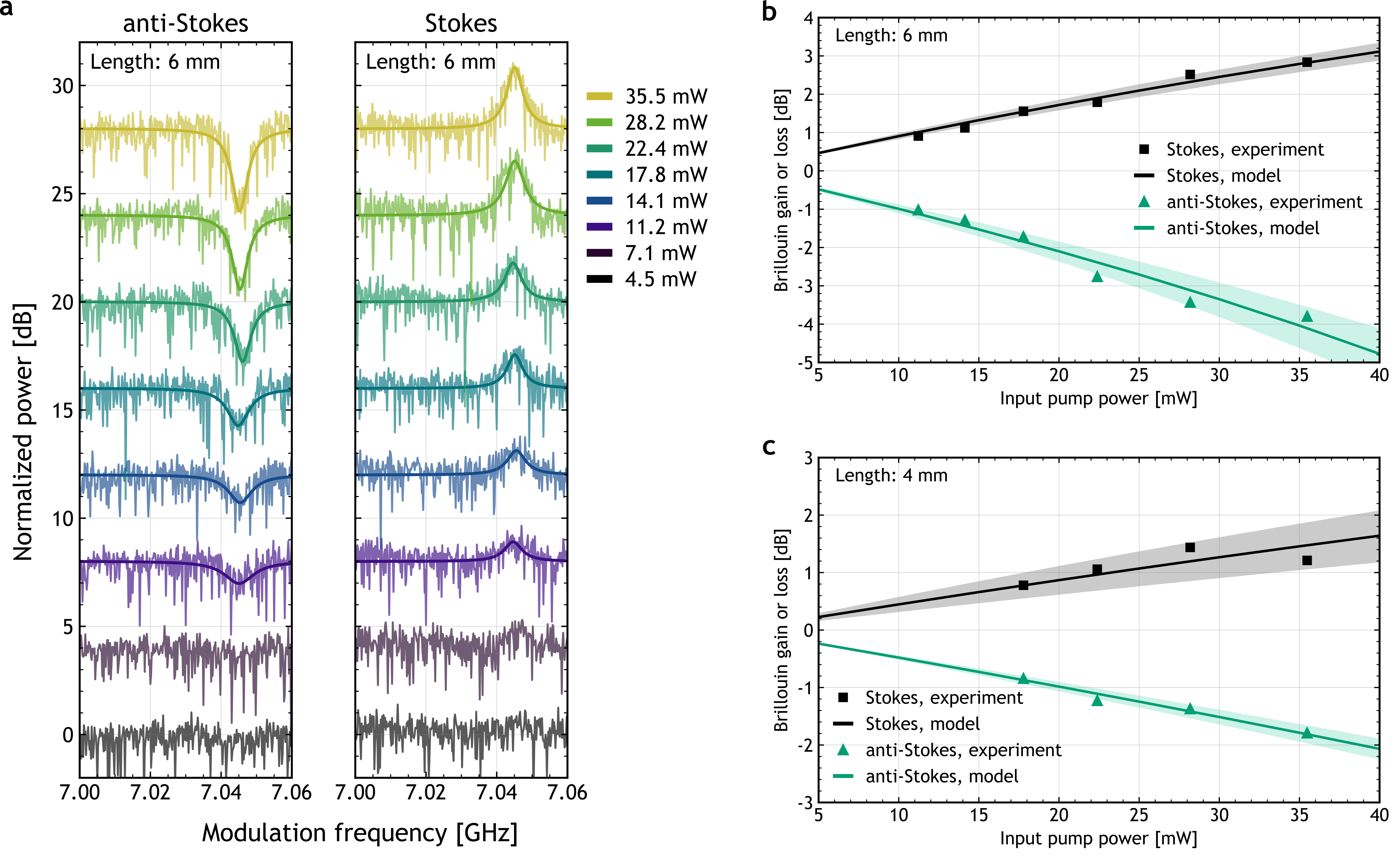}
    \caption{Brillouin gain and loss in the proposed optomechanical waveguide. \textbf{a} Experimental RF beat note and theoretical fitting between the reference line and the Stokes or the anti-Stokes lines as a function of the modulating frequency for different on-chip input pump powers. \textbf{b}, \textbf{c} Brillouin gain (black) and loss (green) of the Stokes and anti-Stokes, respectively, for two different waveguide lengths: 6 mm (\textbf{b}) and 4 mm (\textbf{c}). Here, squares and triangles represent the experimental measurements, while the solid lines stand for the theoretical model. The shaded region highlights the 3$\sigma$ confident interval for the gain coefficient.}
    \label{fig:peak_power}
\end{figure}

In Figure \ref{fig:peak_power}a, we illustrate the normalized intensities of Stokes and anti-Stokes light as a function of the modulating frequency for various input pump powers. In each case, the input powers for Stokes and anti-Stokes probes are set to $\sim28$ \unit{\dB} lower than the pump input power. Light is coupled into and out of the system using grating couplers with an efficiency of 16\%. The waveguide under study has a length of 6 \unit{\mm}, with 6.3 \unit{\dB\per\cm} of linear loss (see Supplementary materials, section A).  

The anti-Stokes line experiences a depletion, while the intensity of the Stokes line increases due to the energy transfer mediated by the Brillouin effect. These spectra exhibit a Lorentzian shape centered at the frequency of the mechanical mode that facilitated the coupling, commonly referred to as the Brillouin shift. We obtain the key parameters that characterize the interaction by fitting the experimental data to the nonlinear theoretical model (see Supplementary materials, section B). From this analysis, we extract a mechanical frequency of $\nu_\mathrm{m} = \Omega_\mathrm{m}/2\pi = 7.0452 \pm 0.0001$ \unit{\GHz}, a linewidth of $\gamma_\mathrm{m} = \Gamma_\mathrm{m}/2\pi = 6.4 \pm 0.5$ \unit{\MHz} $(Q_\mathrm{m} = \nu_\mathrm{m}/\gamma_\mathrm{m} \approx 1050)$, and a Brillouin coefficient of $G_\mathrm{B} = 2673 \pm 67$ \unit{\per\W\per\m} for the optomechanical interaction. The measured acoustic frequency shows good agreement with the predicted one ($\nu_\mathrm{m} = 7.184$ \unit{\GHz}). The experimental Brillouin gain coefficient is larger than the simulated value ($G_\mathrm{B} = 2000$ \unit{\per\W\per\m}). By comparing with COMSOL simulations, we conclude that this mechanical mode corresponds to the third-order breathing mode, whose Brillouin gain coefficient is very sensitive to variations of the duty cycle in the SWG cladding (see Supplementary materials, section D). Hence, fabrication deviations and inhomogeneities of the SWG gap along the waveguide length may account for the differences between the experimental and the simulated value of the optomechanical coupling coefficient.

The power dependency of the Brillouin gain and loss aligns well with the theoretical model, as illustrated in Figures \ref{fig:peak_power}b and \ref{fig:peak_power}c for two different waveguide lengths: 6 mm (b) and 4 mm (c). Brillouin interaction is a cumulative effect; therefore, longer waveguides result in greater Brillouin gain or loss. For the 4 mm-long waveguide, the Brillouin parameters obtained by fitting the corresponding RF spectra are $\nu_\mathrm{m} = 7.0426 \pm 0.0002$ \unit{\GHz}, $\gamma_\mathrm{m} = 7.6 \pm 0.9$ \unit{\MHz} $(Q_\mathrm{m} \approx 930)$, and a Brillouin coefficient of $G_\mathrm{B} = 1932 \pm 81$ \unit{\per\W\per\m}. Additionally, we observe an asymmetry between the gain in the Stokes signal and the depletion of the anti-Stokes signal in the longer waveguide, which is only slightly noticeable in the shorter waveguide. In the 6 mm-long waveguide, a maximum depletion of 4 dB is achieved on the anti-Stokes sideband with an input power of 35.5 mW, while the Stokes line at the same input power records a maximum gain of 3 dB.

Despite the strong optomechanical interaction provided by this waveguide, the moderately high linear losses due to surface roughness hinder the achievement of a net gain regime. While enlarging the waveguide width is a common approach in the literature to address this issue, it would also diminish the Brillouin coupling because the overlap between the acoustic and optical mode profiles would decrease. Therefore, efforts to reduce surface roughness must focus on improving fabrication processes \cite{reduc_surfaceRoughness_Lee_2015, reduc_surfaceRoughness_Bellegarde_2018}. For instance, with the experimental Brillouin gain coefficient of 2673 \unit{\per\W\per\m}measured here, reducing the linear loss down to 4 \unit{\dB\per\cm} would be sufficient to achieve net Brillouin gain in our structures. Linear losses of around 2 and 3 \unit{\dB\per\cm} have been reported in SWG structures in earlier works \cite{Bock_SWG_lowloss_2010}. These results, despite the lack of net gain, are adequate for creating narrow-linewidth notch filters \cite{notch_filter_pedestral_2015, BS_Passive_RFfilter_Liu2019}. Furthermore, the extended profile of the mechanical mode in our waveguides result in highly nonlocal nonlinearities \cite{nonLocalBS_Gertler2020}  which could facilitate the coupling different optical waveguides to provide a coherent control of information \cite{InfoManiputaion_CoupledWG_Kim2021} or shape and RF signal \cite{RFshaping_CoupledWG_Kim2022}.

We examine how the mechanical frequency varies with different waveguide and subwavelength grating ($W_\mathrm{c}$ and $W_\mathrm{swg}$). In Figure \ref{fig:freq-geometries}, we present the normalized RF spectra for the Stokes and anti-Stokes lines across various devices. All waveguides are 6 mm long, with an on-chip input pump power of approximately 28 mW. The input power for the Stokes and anti-Stokes lines is set to $\sim28$ dB lower. We can adjust the mechanical frequency within a range of 5 to 8 \unit{\GHz}, as shown in Table \ref{tab:peaks_geom}. Furthermore, the coupling efficiency, expressed by the Brillouin coefficient $G_\mathrm{B}$, remains high in all cases ($G_\mathrm{B} > 1490$ \unit{\per\W\per\m}). The Brillouin parameters obtained from fitting the experimental data to the theoretical model  (see Supplementary materials, section B) align well with the simulated values for all acoustic modes. In Figure \ref{fig:freq-geometries}, we also illustrate the RF beating note of the anti-Stokes line for a waveguide featuring the same SWG cladding without a phononic crystal. This reference waveguide do not exhibit any optomechanical coupling signatures, underscoring the critical role of the phononic bandgap in confining the mechanical modes. We show a tunability range of 3 GHz with Brillouin gain coefficient and mechanical confinement comparable to the state-of-the-art \cite{shin_tailorable_2013, kittlaus_large_2016, ACS_Photonics_3toneGain} using a single etching step and an all-silicon waveguide. 

\begin{figure}
    \centering
    \includegraphics[width=\linewidth]{Fig_RF_reference_peaks.png}
    \caption{Normalized RF spectra for the Stokes (purple) and anti-Stokes (green) sidebands for different transversal geometries of the studied waveguide and an on-chip input pump power of 28 \unit{\mW} (an offset of +1 (-1) is included for the (anti-)Stokes for clarity). In black, we plot the RF beating note of the anti-Stokes line for waveguides with SWG cladding and without phononic crystal. All measured waveguides were 6 \unit{\mm} long. Notations related to waveguide parameters are defined in Figure \ref{fig:geometry}b.}
    \label{fig:freq-geometries}
\end{figure}

{
\renewcommand{\arraystretch}{1.1}
\begin{table}[hbtp]
    \centering
    \caption{Characteristic Brillouin parameters for different geometries. The waveguide length is set to 6 mm. Here, \textit{Exp.} for experimental measurements, and \textit{Sim.} for simulated values using the measured $Q_\mathrm{m}$ for calculations and assuming an optical mode of frequency $\omega_\mathrm{p}/2\pi=193$ THz. Notations related to waveguide parameters are defined in Figure \ref{fig:geometry}b.}
    \label{tab:peaks_geom}
    \begin{tabular}{@{}*{11}{c}@{}}
    \toprule
    & & && \multicolumn{3}{c}{$W_\mathrm{swg}=2$ \unit{\um}} && \multicolumn{3}{c}{$W_\mathrm{swg}=2.25$ \unit{\um}} \\
    \cmidrule{5-7} \cmidrule{9-11}
    & & && $W_\mathrm{c}=600$ nm & \multicolumn{2}{c}{$W_\mathrm{c}=700$ nm} && \multicolumn{2}{c}{$W_\mathrm{c}=600$ nm} & $W_\mathrm{c}=700$ nm \\
    \midrule
    \multirow{4}{*}{\begin{sideways} Exp. \end{sideways}} & $\nu_\mathrm{m}$ & \unit{\GHz} && 7.0452 & 5.4278 & 6.7630 && 6.5044 & 7.7658 & 6.2614 \\
    & $\gamma_\mathrm{m}$ & \unit{\MHz} && $6.4\pm0.5$ & $4.5\pm0.2$ & $6.8\pm0.4$ && $6.3\pm0.7$ & $7.0\pm0.4$ & $4.6\pm0.3$ \\
    & $Q$ & -- && 1050 & 1200 & 990 && 1050 & 1100 & 1365 \\
    & $G_\mathrm{B}$ & \unit{\per\W\per\m} && $2673\pm67$ & $1926\pm87$ & $1490\pm65$ && $2461\pm66$ & $1850\pm136$ & $1866\pm163$ \\
    \midrule
    \multirow{2}{*}{\begin{sideways} Sim. \end{sideways}} & $\nu_\mathrm{m}$ & \unit{\GHz} && 7.156 & 5.5087 & 6.7630 && 6.6037 & 7.8872 & 6.3629 \\
    & $G_\mathrm{B}$ & \unit{\per\W\per\m} && 2000 & 1965 & 1318 && 1457 & 1485 & 2109 \\
    \bottomrule
    \end{tabular}
\end{table}
}

\section{Conclusion}
We propose and demonstrate a new design strategy for optomechanical waveguides that utilizes periodic silicon nanostructures to achieve efficient Brillouin interactions. Our design incorporates a phononic crystal to confine mechanical modes, along with a subwavelength cladding to facilitate optical propagation with moderate losses. Notably, this design requires only a single etching step, which simplifies the fabrication process. We have achieved state-of-the-art optomechanical coupling, with an experimental gain coefficient of $G_\mathrm{B}$ up to 1380 \unit{\per\W\per\m}. Additionally, we measured a Stokes gain of 1.5 dB and an anti-Stokes depletion of $-2$ dB using a 6 mm-long waveguide with an on-chip input power of 35 mW. Our design allows for seamless selection of the mechanical frequency within the range between 5 and 8 \unit{\GHz} with comparable performance by simple geometry modifications. To our knowledge, this work represents the first experimental demonstration of subwavelength engineering applied to optomechanical waveguides. These findings open interesting perspectives for implementing narrow linewidth RF notch filters, studying of nonlocal nonlinearities in optical systems, and combining optomechanical interactions with dispersion and anisotropy engineering thanks to the degrees of freedom of the subwavelegth cladding \cite{cheben_subwavelength_2018,Cheben_MetamaterialsIntegrated_2023}. 

\medskip
\textbf{Supporting Information} \par 
Supporting Information is available from the Wiley Online Library or from the author. 

\medskip
\textbf{Acknowledgements} \par 
The authors thank the Agence Nationale de la Recherche for supporting this work through BRIGHT ANR-18-CE24-0023-01 and MIRSPEC ANR-17-CE09-0041. This project has received funding from the European Union's Horizon Europe research and innovation program under the Marie Sklodowska-Curie grant agreement Nº 101062518. This work was partially funded by the European Union through the European Research Council (ERC) project SPRING (Grant agreement No. 101087901). This publication is part of the grant RYC2023-045670-I funded by MICIU/AEI/10.13039/501100011033 and by ESF+. This work was done within the C2N micro-nanotechnologies platforms, and partly supported by the RENATECH network and the General Council of Essonne. We acknowledge Xavier Le Roux for his significant contributions to the development of the fabrication processes used in this study. 

\medskip
\textbf{Conflicts of interest} \par
The authors declare no conflict of interest. We disclosure the use of the educational tool Grammarly (and its AI Writing Support) for proofreading (grammar and spelling checking) and for punctual clarity improvements during the preparation of the manuscript.


\medskip

\bibliographystyle{MSP}
\bibliography{references}


\newpage
\textbf{Table of contents} \par


\end{document}